\documentstyle{report}
\begin{document}
\begin{center}
{ \Large Retarded-Advanced N-point Green Functions \\ in thermal
field theories }
\\
06108 Nice Cedex 2, France}
\end{center}

\bigskip
\centerline{\large \bf Abstract}

\bigskip
   The general relationship between a Retarded-Advanced Green
function of the Real Time formalism and analytical
continuations of the Imaginary Time Amplitude
 is obtained via perturbation theory. A 4-point R-A
function is, at most, a linear combination of three analytical
continuations, and a 5-point R-A function of seven continuations.
The R-A functions are written as the sum of one analytical
continuation plus discontinuities  across
single momentum transfer variables. These discontinuities
correspond to processes where particles are on
shell in an intermediate state with a thermal statistical
weight.

For the 3-point function, the difference between the real
parts of two different analytical continuations is a
double discontinuity. This is expressed in terms of
three one-particle decay amplitudes.

Feynman rules in the R-A basis introduce a new feature,

the flow of  infinitesimal $\epsilon$ , which obeys
Kirchoff laws.
\vfill
\begin{flushright}
INLN 1993/14 \ \  May 93
\end{flushright}

\newpage
 \begin{center}
{\Large \bf 1.  \ Introduction}
\end{center}

\bigskip

\noindent Recently in finite temperature field theory, there has
 been an increased interest in 3-point  and N-point
functions. Using the path integral approach to the Real Time
formalism and using spectral representations, Evans [1] has shown
that there exist two general linear relations

  i) between the Real-Time Green
functions (RT), expressed in terms of doublet fields, and the
Retarded-Advanced functions (RA)

ii) between the Retarded-Advanced functions
 and  analytical continuations of the Imaginary-Time
formalism amplitudes (IT).
\\  The proof is valid quite generally for
an N-point function and it does not depend on the details of the
theory nor on the statistics of the fields. Evans showed the

explicit relations for the 3-point function, which generalized
previous results [2,3].

Independently, Aurenche and Becherrawy [4] presented a
reformulation of the RT perturbation theory in terms of retarded
(R) and advanced (A) propagators, instead of using the usual
doublet field (1,2). They determined the U,V matrices that allow
to pass from the (1,2) basis to the (R,A) basis. This allowed a

formulation of Feynman rules in the (R,A) basis and they gave one
and two-loop examples of RA Green functions for bosons and
N=2,3,4-point functions. Van Eijck and Van Weert [5] generalized
those Feynman rules to any interaction and both statistics. Taylor
[6] used the U matrix of ref.[4] and the spectral representation
of ref.[1] to investigate the relation between RA and IT
amplitudes for the 4-point function.

\bigskip
Evans [1] has noted that for $N>3$, the situation is complicated.
 He has shown that the useful analytical continuations in the
external energy variables from discrete imaginary towards the real
axes are

$i\nu_j\ 2\pi T \ \rightarrow \ p_j+i\epsilon_j$
 \\
where the infinitesimal $\epsilon_j$ may be positive or
negative. He observes that the result of an IT calculation
depends not only on the signs of the individual $\epsilon_j$
but also on the signs of all the possible sums of subsets of

$\epsilon_j$. For $N>3$, the constraint $\sum_{j=1}^N

\epsilon_j = 0 $ is not enough to determine all these sums.  The
choice of an RA amplitude fixes the individual $\epsilon_j$.  Once
they are chosen, different analytical continuations of the IT
amplitude correspond to different signs for the undetermined
sums. In this case the RA amplitude is a linear combination of
those continuations.

\medskip
The relation between RA and IT amplitudes is quite general. In this
paper, diagrams in perturbation theory will be used to find it

for $N>3$-point Functions. Compared to the work in ref.[1,6] an
added input will be the analytical properties of an IT amplitude.
These properties are explicit in the forms presented in ref.[7] for
an IT amplitude associated to a Feynman diagram.

\noindent The RA and IT amplitudes will be compared at the tree and
loop level. One then will obtain the RA amplitude as the sum of one
analytical continuation plus discontinuities in specific "crossed"
channels.  These discontinuities will be associated to process
where particles are on shell in an intermediate state. The
intermediate states are the familiar ones at $T=0$, with a thermal
statistical weight. This work will be restricted to bosons, and as
shown in ref.[1,5], in the case of bosons and fermions the
coefficients are obtained with a simple substitution.

An interesting byproduct is found, the general form
for the difference between the real parts of two different
analytical continuations of the 3-point function. That form
generalizes the result obtained at the one-loop level [8], a
result in connection with the definition of an effective coupling

constant.

\bigskip
In section 2 is a review of one form for the IT amplitude
associated to a diagram. The Feynman rules are reviewed in the R,A
basis and  an useful tool is introduced, the flow of

infinitesimal epsilon.  Those rules for individual
diagrams are compared to the general relation between RA and IT
amplitudes  in the simplest case.

In sect. 3 the important general relations between RA and IT
amplitudes are written explicitely for $N=4,5$,  then generalized.

In sect. 4  the discontinuity in an individual momentum-transfer
energy-variable  is shown to  be a sum of terms. These terms

factorize into two IT amplitudes and a thermal statistical weight
associated to the multiparticle intermediate state in that channel
. The proof will use the $\epsilon$-flow. With the same method,
the difference between the real parts of two analytical
continuations of the 3-point function will be obtained in
sect.5. Conclusions will be drawn in in sect.6.

\bigskip

\bigskip

\begin{center}
{\Large \bf 2.  \  Feynman Rules and Examples}
\end{center}
\setcounter{chapter}{2}

\bigskip

\bigskip

{\large \bf 2.1  \ The Imaginary Time Amplitudes}

\bigskip
The statement is restricted to the case of bosons. The
property of an IT amplitude to be used repeatedly follows.

 For an n-loop diagram contributing to an N-point
function, the integrand may be written as the sum over all
possible trees obtained by cutting n internal lines so that a
connected tree diagram is obtained [7,4].

  i) To a cut internal
line is attached the factor

\begin{equation}
A_i=\delta({\underline{k}}_i^2-m^2)\epsilon(k_i^o)(n(k_i^o)+1/2)
\end{equation}
\noindent where the energy $k_i^0$ is real, $n(k_i^0)$ is
the Bose-Einstein factor, and $\epsilon(k_i^0)=\pm1$ according

to the sign of $k_i^0$ .

ii) The external energies only enter
the propagators associated to the tree's internal lines.

\medskip
 The chain of arguments leading to that property is briefly
recalled [7]. In the Imaginary Time formalism, all energies are
initially discrete imaginary. Keeping external momenta discrete,
one may perform the summations over the energies running around the
loops, in successive order, with the use of

\begin{equation}
T\sum_n f(z_n=in2\pi T) =\int_C{{{\rm
d}z\over{2i\pi}}f(z){1\over2}\coth{\beta\over2}z}
\end{equation}
\noindent where the contour $C$ runs initially around the poles
of the coth function and is deformed into the complex $z$ plane
to pick up all the poles $z_i$ of $f(z)$. In a Feynman diagram,
the poles come from the internal lines' propagators. To pick up
a  pole $z_i$ is to put an internal line $i$ on shell. The
result is equivalent to multiply the integrand by a factor
$$\int_{-\infty}^{+\infty}{\rm d}k_i^0 \epsilon(k_i^0)
\delta((k_i^0)^2-{\bf k}_i^2-m^2)
{1\over2}\coth{\beta\over2}k_i^0$$
\noindent With
${1\over2}\coth{\beta\over2}k_i^0=n(k_i^0)+1/2$, one obtains
the form (2.1) for $A_i$. After performing the n summations, one
analytically continue the external energies variables $p_j^0$
from discrete imaginary towards real values $p_j^0\pm
i\epsilon_j$. As a result, the trees' propagators develop poles.
These determine the analytical properties of the

amplitude as a function of those external energies.

For an n-loop diagram, the above rule does not quite give the
full integrand. For each tree, a term with the product of n
factors $A_i$ is certainly present, as one may perform the n
summations in any order. As shown in ref.[7], for some trees,
additional terms should appear in the numerator, which are lower
 order terms in
powers of the B.E. factor. Rules were given in ref.[7] to generate
them up to the fourth loop order. The exact form of the trees'
 numerators is irrelevant to the amplitude's analytical
properties that  will be used in this section.

\bigskip

{\large \bf 2.2  \ Feynman Rules and Epsilon-flow
for RA Amplitudes}

\bigskip

The Feynman rules for diagrams in the R,A basis for bosons in a
$\Phi^3$ theory were set up in ref.[4]. The retarded and advanced
propagators are
\begin{equation}
\Delta_R(\underline{k})=
\Delta_R(k_0,{\bf k})={1\over{(k_0+i\epsilon)^2-{\bf k}^2-m^2}}
=\Delta_A^*(k_0,{\bf k})
\end{equation}
\begin{equation}
\Delta_R(\underline{k})-\Delta_A(\underline{k})=
2\pi\epsilon(k_0)\delta({\underline{k}}^2-m^2)
\end{equation}
\begin{equation}
\Delta_A(k_0,{\bf k})=\Delta_R(-k_0,-{\bf k})
\end{equation}
 There are two types of vertices; with all momenta incoming
\begin{equation}
V_{RRA}^{pqr}=g \ \ \ ,\ \ \ V_{AAR}^{pqr}=-N (p,q)\
V_{RRA}^{pqr\  *}
\ \ \ , \ \  V_{RRR}=0=V_{AAA}
\end{equation}
\begin{equation}
N (p,q) = {n(p_0) n(q_0)\over{n(p_0+q_0)}}
\end{equation}
where $n(p_0)$ is the Bose-Einstein factor. Useful properties
are
\begin{equation}
N (p,q) =1 + n(p_0) + n(q_0)
\end{equation}

\begin{equation}
1 + n(p_0) + n(-p_0) = 0
\end{equation}

\begin{equation}
N(p,q) N(p+q,r)={n(p_0)n(q_0)n(r_0)\over{n(p_0+q_0+r_0)}}
\end{equation}

\begin{equation}{\prod_{i=1}^N
n(q_i^0)\over{n(\sum_{i=1}^Nq_i^0)}}=
\prod_{i=1}^N(1+n(q_i^0))-\prod_{i=1}^Nn(q_i^0)={\cal N}(q_1,q_2,
\ldots,q_N)
\end{equation}
This last equation is proved by recurrence. It is valid for
$n(q_1^0+q_2^0)$, one then considers $N(q_3,q_2'=q_1+q_2)$
with (2.7) and (2.8). The factors $N(p,q)$ and
${\cal N}(q_1,q_2,\ldots,q_N)$ will play a central role in what
follows. To generalize forms (2.8) and (2.11) to the case
of mixed fermions and bosons is simple, $n(q_i^0)$ is replaced by
$\eta_in(q_i^0-\mu_i)$ where, for fermions, $\eta_i=-1$ and
$\mu_i$ is the chemical potential, see ref.[5].

 The main feature of those rules is, a vertex with a pair
of A legs is weighted by the factor $N(p,q)$ . A recurrent
feature of the R,A basis will turn out to be, N legs A are
weighted by the factor ${\cal N}(q_1,q_2,\ldots,q_N)$. Note that,
with another choice of the U,V matrices between the (1,2)
and (R,A) basis, the weight would be on the R legs, and
none would be on the A legs [4].

\bigskip
 Graphical rules are now introduced. One may consider the
R,A components of the propagator as the analogue of the two
components of a charged scalar field, and one may keep
track of both components with an arrow on the internal lines. An
R line with momentum $(-k_0,-{\bf k})$ is equivalent to an
A line with momentum $(k_0,{\bf k})$, from (2.5) .
The bare vertices are drawn on fig.1. That new type of
arrow describes the oriented flow of  infinitesimal
epsilons. This flow obey Kirchoff laws, as the vertex'
momenta are such that $p+q+r=0$ and
$\epsilon_p+\epsilon_q+\epsilon_r=0$ for the analytical
continuation $p+i\epsilon_p, q+i\epsilon_q, r+i\epsilon_q$

\noindent For an N-point Green function, the conventions will be

all external momenta are incoming

an external line with an
$\epsilon$-arrow incoming (resp. outgoing) the Green function is
an R line (resp. A line).

\\
If  both orientations of an internal line are
allowed at a vertex, one considers  the two cases for the flow.
One sums over all diagrams corresponding to all possible
cases for the flow along the internal lines, from the
external R lines (sources) towards the external A lines
(sinks). The graphical rules of fig.1 extend to the
case of dressed propagators and vertices (see later on
(2.17) and (2.16) )

\bigskip

{\large \bf 2.3  \  Green functions with one A or one R}

\bigskip

 The tree diagrams contributing to a Green
function $\Gamma_{AR\ldots R}^{pq_1\ldots q_N}$ are considered.

In terms of the $\epsilon$-flow, there are one sink and
many sources and the flow along the tree's propagators
is fixed in an unique way. All vertices are of the RRA
type and the amplitude is related to one analytic
continuation of the IT amplitude
\begin{equation}
\Gamma_{AR\ldots R}^{pq_1\ldots
q_N}=\Gamma^{N+1}(p_0+i\epsilon_p,
q_1+i\epsilon_1,\ldots,q_N+i\epsilon_N)
\end{equation}
with $\epsilon_p<0\ ,\ \ \epsilon_i>0\ i=1,\ldots,N$
and  $\epsilon_p+\sum_i\epsilon_i=0$. That relation
has been found at the tree level. From Evans' proof the
relation is valid quite generally and  was given in ref.[1]. The
implications for an $n$-loop diagram are now described.

As recalled in sect.2.1, the IT amplitude has the following
property, the integrand may be written as the sum over all the
possible tree diagrams that are obtained by cutting $n$ internal
lines so that a connected tree diagram results. If one adopts the
rule

a cut line is a dead end for the flow

\\
then the $\epsilon$-flow along the internal lines of each tree is
fixed in an unique way, as there is only one sink. As a result,
the right-hand side of form (2.12) is written as a sum of those
oriented trees, and an explicit form for the RA amplitude is
obtained.

 Starting fron the RA Feynman rules given in sect.2.2,
$\Gamma_{ARR}$ was obtained in ref.[4] as a sum of oriented trees,
at the one-loop and two-loop level, and relation (2.12) was
found in those cases. In the appendix, a one-loop example shows
how the sum of oriented trees arises from the sum over all possible
cases for the $\epsilon$-flow around the loop, thanks to (2.4).

\bigskip

The Green functions with only one R leg are now considered. For

 a tree diagram contributing to
$\Gamma_{RA\ldots A}^{pq_1\ldots q_N}$, the $\epsilon$-flow is
fixed along the tree's internal lines, as there is only one
source. Each vertex is of the  RAA type and carries a factor $N$.
The amplitude is related to one analytic continuation of the IT
amplitude
\begin{equation}
\Gamma_{RA\ldots A}^{pq_1\ldots q_N}=\Gamma^{N+1}
(p_0+i\epsilon_p,q_1+i\epsilon_1,\ldots,q_N+i\epsilon_N)(-1)^{N-1}
{\cal N}(q_1,q_2,\ldots,q_N) \end{equation}
with $\epsilon_p>0\ ,\ \ \epsilon_i<0\ i=1,\ldots,N$ and
$\epsilon_p+\sum_i\epsilon_i=0$ , where
\begin{equation}
{\cal N}(q_1,q_2,\ldots,q_N)
=N(q_1,\sum_{j\neq1}q_j)N(q_2,\sum_{j\neq
1,2}q_j)\cdots N(q_{N-1},q_N) \end{equation}
Repeated use of (2.10) leads from form (2.14) to form (2.11)

for ${\cal N}(q_1,q_2,\ldots,q_N)$. Relation (2.13) is valid
 quite generally. For an $n$-loop diagram the IT amplitude

 is made of trees where  branches are stuck along a
genuine tree diagram. Those dead ends do not modify the direction
of the flow, and one learns that the cut-lines' energies dont show
up in the relation (2.13) between the RA and IT amplitudes. An
explicit one-loop example for $\Gamma_{RAA}$ is eq.(4.24) in
ref.[4].

\\  Comparing forms (2.12) and (2.13), one obtains
\begin{equation}
\Gamma_{RA\ldots A}^{pq_1\ldots q_N}=\Gamma_{AR\ldots R}
^{pq_1\ldots q_N\  \ast}\ (-1)^{N-1}\ {\cal N}(q_1,q_2,\ldots,q_N)
\end{equation}
Relation (2.15) was obtained earlier in ref.[5] for the particular
case of bare vertices. The authors used the (U,V) matrices

to go from the (1,2) basis to the (R,A) basis in a theory with
multiparticle bare vertices in the RT formalism.

\medskip
Special cases of relation (2.15) are the dressed vertex, and the
self-energy [4]

\begin{equation}
\Gamma_{AAR}^{pqr}=
\Gamma_{RRA}^{pqr\ \ast}\ (-1) \ N(p,q)

\end{equation}
\begin{equation}
\Gamma_{R\ \ A}^{p,-p}=\Gamma_{A\ \ R}^{p,-p\
\ast}=\Sigma(p_0+i\epsilon,|{\bf p}|)
\end{equation}
where $\Sigma$ is the IT amplitude. Since $\Sigma$ is an even
function of $p_0$, relation (2.5) holds for the dressed
propagator and the diagrammatic rules of fig.1 are valid for
the dressed case.

\bigskip
In this section we have shown that the double degree of
freedom  takes a simple form in the (R,A) basis,
the oriented flow of  infinitesimal epsilons. The general relations
between RA and IT N-point amplitudes for the case only one A, or
only one R
 are (2.12) and (2.13).

\bigskip

\bigskip

\begin{center}
{\Large \bf 3. \

Green Functions with more than one A (or R)}
\setcounter{chapter}{3}
\setcounter{equation}{0}
\end{center}

\bigskip

\bigskip
For the amplitudes with more than one A (or R) i.e. more than
one sink (or source), the $\epsilon$-flow along the tree's
propagators is not completely determined, and one has to sum
over all allowed directions. Different orientations of the flow
will correspond to different ways of approaching the real axis
in some energy variable, with a characteristic weight factor.
The relations between the RA and IT amplitudes are more
complicated. The  4-point and
5-point functions will be examined successively, then general
statements will be made.

\bigskip

{\large \bf  3.1 \ The 4-point function $\Gamma_{AARR}^{\ pqsr} $}

\bigskip
 The relevant analytical continuations of the energy
variables, $p_0+i\epsilon_p\ ,\ q_0+i\epsilon_q\ ,\
r_0+i\epsilon_r\ ,\ s_0+i\epsilon_s$  are such that

$\epsilon_p+\epsilon_q+\epsilon_r+\epsilon_s=0$ with
$\epsilon_p,\epsilon_q <0$ and $\epsilon_r,\epsilon_s  >0$.
\\
If one considers the two-particle energy variables, both signs
are allowed for $\epsilon_p+\epsilon_r$ and
$\epsilon_p+\epsilon_s$ . They shall be referred  to  as
$(p+r)_R,(p+r)_A,(p+s)_R,(p+s)_A$.

First, the tree diagrams of fig.2 are considered. One has to sum
over both orientations of the flow along the lines carrying the
momentum $p+r$ or $p+s$ and there is one vertex of the type RAA.
With the Feynman rules of fig.1 one obtains \begin{eqnarray}
- \ \Gamma_{AARR}^{\ pqsr}&=&N(p,q+s) \ F_1((p+r)_R) +
N(q,p+r) \ F_1((p+r)_A)  \nonumber\\
& & \mbox{}+N(p,q+r) \ F_2((p+s)_R) + N(q,p+s) \ F_2((p+s)_A)
\nonumber\\ & & \mbox{}+ N(p,q) \ F_3((p+q)_A)
\end{eqnarray}
while the IT amplitude is
\begin{equation}
\Gamma^4(p+r,p+s,p+q)=F_1(p+r)+F_2(p+s)+F_3(p+q)
\end{equation}
The coefficients obey the sum rule
\begin{equation}
N(p,q+s)+N(q,p+r)=N(p,q)=N(p,q+r)+N(q,p+s)
\end{equation}
since, from (2.9), $1+n(q+s)+n(p+r)=0$ for $p+q+r+s=0$. From
(3.1) (3.2) (3.3) one gets
\begin{eqnarray}
-\ \Gamma_{AARR}^{
\ pqsr}&=&\Gamma^4((p+r)_A,(p+s)_A) \ \ N(p,q)\nonumber\\ &
&\mbox{}+N(p,q+s) \ [\Gamma^4((p+r)_R,(p+s)_A)-
\Gamma^4((p+r)_A,(p+s)_A)] \nonumber\\ &
&\mbox{}+N(p,q+r)
\ [\Gamma^4((p+r)_A,(p+s)_R)-\Gamma^4((p+r)_A,(p+s)_A)]
\end{eqnarray}

$\Gamma_{AARR}^{\ pqsr}$ is a linear combination of
 three analytical continuations of the IT amplitude. The choice
of $(p+r)_A,(p+s)_A$ for the first term is arbitrary. In the
second term only $F_1((p+r)_R)-F_1((p+r)_A)$ survives.

\noindent The relation (3.4) is the looked-for general relation,
where the IT amplitude $\Gamma^4$ also depends on the energy
variables $p_A,q_A,r_R,s_R,(p+q)_A$. Forms (3.1) (3.2) (3.3)
are valid for any multiloop diagram as it is now described. The
IT amplitude associated to a loop diagram can be written as a sum
of trees; as each tree depends only on $one$ two-particle energy
variable, the sum of trees split into three classes, and the
IT amplitude may be written as form (3.2). For the
$\Gamma_{AARR}$ amplitude, each tree that depends on $p+r$ occurs
twice, one for each orientation of the flow along the lines
carrying the momentum $p+r$; $(p+r)_R$ occurs with the factor
$N(p,q+s)$, $(p+r)_A$ with the factor $N(q,p+r)$.

\\
As an example the trees
associated with the one-loop diagram are shown on fig.3. The
IT amplitude has form (3.2) with $F_2(p+s)=0$. For
 $\Gamma_{AARR}$, one sums over both orientations of the flow along
$p+r$ in diagrams (a) and (b); the amplitude was computed in
ref.[4] from the RA Feynman rules and obtained as a sum of trees,

eq.(4.49) of ref.[4] is form (3.1) for that case. In the appendix,
this one-loop example shows how the RA Feynman rules lead to a sum
of oriented trees  with the extra  factors $N$ .

\bigskip
The general form (3.1) is drawn on fig.4 in a diagrammatic way
which emphasizes the $\epsilon$-flow. The double line is for the
tree's internal lines carrying the momentum $p+r$ and the blobs
indicate the tree's parts which depend on the individual energy
variables $p,r,q,s$. The important feature is, the flow is uniquely
determined in each blob. For diagram (a) the left blob has a
single source, the right one a single sink. All four terms of
fig.4 split the $\epsilon$-flow in the same way, one source-one
sink. The term $F_3$ in form (3.1) may be included in $F_1$ or in
$F_2$ with the use of (3.3).

 Some physical interpretation may be given to the results.
In form (3.4) one analytical continuation of the

amplitude has been singled out and  the two other terms are
discontinuities in one energy variable $p+r$ (or $p+s$).
The following properties hold (see sect.4):

i) such a
discontinuity corresponds to processes where particles are
on shell in one of the possible intermediate states of the
amplitude in the $p+r$ (or $p+s$) channel

ii) in an IT
amplitude, a multiparticle intermediate state
$(q_1,\ldots,q_m)$ carries the weight ${\cal N}(q_1,\ldots,q_m)$
as given by (2.11).  \\
Then,  the factors arising in
(3.1) and (3.4) may be interpreted as follows. A factor
$$N(p,q+s)={n(p)n(q+s=\sum_{i=1}^mq_i)\over{n(p+q+s)}}$$
transforms the weight so that it is symmetric with respect
to $p$ and to $q_1,\ldots,q_m$ , i.e. with respect to all
the momenta carrying an out-flowing $\epsilon$ in the
intermediate state. This is  the case for  the first term of form
(3.1) and for diagram (a) on fig.4.

\noindent If one now imagines that the amplitude
$\Gamma_{AARR}^{\ pqrs}$ is associated to a collision

$$r\ \ +\ \ s\ \ \rightarrow\ \ p\ \ +\ \ q$$
$p+q$ is the direct channel, $p+r$ and $p+s$ the crossed
channels. Form (3.4) accounts for the occurrence of processes
where intermediate particles are on shell in a crossed channel.
At $T=0$ kinematics forbids such processes in the physical region
of the direct channel; at $T\not=0$, those processes are a
manifestation of the plasma, particles $r$ and $s$ interact via
the plasma.

\bigskip

{\large \bf 3.2 \ Five-point Function}

\bigskip

\noindent{  \bf 3.2.1 The amplitude $\Gamma_{AARRR}^{ \
pqrst}$}

\bigskip
\noindent The analytical continuations of the energy variables are
such that

$\epsilon_p+\epsilon_q+\epsilon_r+\epsilon_s+\epsilon_t=0$ with
$\epsilon_p,\epsilon_q<0$ and
$\epsilon_r,\epsilon_s,\epsilon_t>0$.
  \\

The convenient variables, whose analytical continuation
is undetermined, are the two-particle subenergies $p+r\ ,\ p+s\ ,\
p+t\ ,\ q+r\ ,\ q+s\ ,\ q+t$ . The decomposition in terms of the
$\epsilon$-flow produces two types of splitting, shown on fig.5 :
(one sink-one source) for diagrams of type (a), (one source-one
sink-one source) for diagrams of type (b). In each blob the flow
is uniquely determined. Trees can only depend on two of the
two-particle subenergies. One considers the trees which depend on
the pair $p+r\ ,q+s$ ; if the associated IT amplitude is
$F(p+r,q+s)$ , the RA amplitude is

\begin{eqnarray}
-\ \Gamma_1&=&N(p,q+s+t) \ F((p+r)_R,(q+s)_A)+
N(q,p+r+t) \ F((p+r)_A,(q+s)_R)
 \nonumber\\
& &\mbox{}+N(p+r,q+s) \ F((p+r)_A,(q+s)_A)
\end{eqnarray}
where the last term corresponds to diagram (b) on fig.5 and the
others to  diagrams of type (a). One does not obtain
$((p+r)_R,(q+s)_R)$. Indeed if

$\epsilon_p+\epsilon_r>0\ \ ,\ \ \

\epsilon_q+\epsilon_s=-(\epsilon_p+\epsilon_r+\epsilon_s)<0 $ .
 \\

 The coefficients obey the sum rule
\begin{equation}
N(p,q+s+t)+N(p+r,q+s)+N(q,p+r+t)=N(p,q)
\end{equation}
since $n(q+s+t)+n(p+r)+1=0$. Form (3.5) may be written
\begin{eqnarray}
- \ \Gamma_1&=&N(p,q) \ \ F((p+r)_A,(q+s)_A) \nonumber \\ & &
\mbox{}+N(p,q+s+t) \ [F(p+r)_R,(q+s)_A)-F((p+r)_A,(q+s)_A)]
\nonumber \\ & & \mbox{}
+N(q,p+r+t) \ [F((p+r)_A,(q+s)_R)-F((p+r)_A,(q+s)_A)]
\end{eqnarray}
There are similar relations for the five other possible pairs
of subenergies. There are also trees which depend on one (or
none) of those subenergies. For example the trees depending on
$(p+q)$ carry the factor $N(p,q)$ . More generally, they are the
analogue of $F_3$ in (3.1) and are included in the blobs on fig.5
in the same way. From (3.7) one obtains the general relation that
links one RA amplitude to seven analytic continuations of the IT
amplitude $\Gamma^5$
\begin{eqnarray}
- \ \Gamma_{AARRR}^{\ pqrst}&=&N(p,q)
\ \ \Gamma^5((p+r)_A,(p+s)_A,(p+t)_A, (q+r)_A,(q+s)_A,(q+t)_A)
\nonumber \\ & & \mbox{}+N(p,q+s+t)\ [\Gamma^5((p+r)_R,others A)-
\Gamma^5((p+r)_A,others A)]
\nonumber \\ & & \mbox{}
+N(p,q+r+s) \ [\Gamma^5((p+t)_R,oth.A)-\Gamma^5((p+t)_A,oth.A)]
 \nonumber \\  & &\mbox{}
+N(q,p+r+t) \ [\Gamma^5((q+s)_R,oth.A)-\Gamma^5((q+s)_A,oth.A)]
\nonumber \\ & & \mbox{}
+N(p,q+r+t) \ [\Gamma^5((p+s)_R,oth.A)-\Gamma^5((p+s)_A,oth.A)]
\nonumber \\ & & \mbox{}
+N(q,p+s+r) \ [\Gamma^5((q+t)_R,oth.A)-\Gamma^5((q+t)_A,oth.A)]

\nonumber \\ & & \mbox{}
+N(q,p+s+t) \ [\Gamma^5((q+r)_R,oth.A)-\Gamma^5((q+r)_A,oth.A)]
\end{eqnarray}

$\Gamma^5$ also depends on the energy variables
$p_A,q_A,r_R,s_R,t_R,p_A+q_A,r_R+s_R,r_R+t_R,s_R+t_R$. In (3.8)
the first term has the statistical weight of the momenta $p$ and
$q$ that carry the outflowing $\epsilon$, just as in (3.4) and
(2.13). The six other terms are the discontinuity in one
individual subenergy variable and only the trees that depend on
that subenergy contribute. As will be shown in sect.4, the first
discontinuity puts on shell all particles of a possible
intermediate state in the $(p+r)$ channel. Then the factor
$N(p,q+s+t)$ transforms the weight so that it is symmetric with
respect to $p$ and to those intermediate particles.

\bigskip

\noindent{  \bf  3.2.2 \ The amplitude $\Gamma_{RRAAA}^{\
pqrst}$}

\bigskip
 On fig.5, all orientations should be reversed. For the
pair of variables $p+r,q+s$ the analogue of (3.5) is

\begin{eqnarray}
\Gamma_2&=&N(r,q+s+t)N(t,q+s) \ F((p+r)_R,(q+s)_A)
\nonumber \\ & & \mbox{}
+N(s,p+r+t)N(t,p+r)
 \ F((p+r)_A,(q+s)_R) \nonumber \\ & &

\mbox{}+N(r,q+s+t)N(s,p+r+t) \ F((p+r)_R,(q+s)_R)
\end{eqnarray}
as there are now two vertices of the RAA type in a tree diagram.
The coefficients obey the sum rule
\begin{eqnarray}
\lefteqn{
N(r,q+s+t)N(t,q+s)+N(s,p+r+t)N(t,p+r)}
\nonumber \\ && \mbox{}
+N(r,q+s+t)N(s,p+r+t)= N(s,t)N(s+t,r)
\end{eqnarray}
with $N(s,t)N(s+t,r)={\cal N}(s,t,r)$ from (2.10) and (2.11).
Relation (3.10) may be proved with the help of

 \begin{equation}
N(s,p+r+t)+N(t,q+s)=N(s,t)
\end{equation}
\begin{equation}
N(s,p+r+t)N(t,p+r)=N(s,t)N(s+t,p+r)
\end{equation} Form (3.9) may be written

\begin{eqnarray}
\Gamma_2&=&{n(s)n(t)n(r)\over{n(s+t+r)}}F((p+r)_R,(q+s)_R)
\nonumber \\  & & \mbox{}+{n(r)n(t)n(q+s)\over{n(r+t+q+s)}}
[F((p+r)_R,(q+s)_A)-F((p+r)_R,(q+s)_R)] \nonumber \\  & &
\mbox{}+{n(t)n(s)n(p+r)\over{n(t+s+p+r)}}
[F((p+r)_A,(q+s)_R)-F((p+r)_R,(q+s)_R))] \end{eqnarray}
 (3.13) is similar to (3.7) and   the general
relation for $\Gamma_{RRAAA}$ is similar to (3.8). The first term
has the statistical weight associated to the three particles
carrying an outflowing $\epsilon$, just as $\Gamma_{RAAA}$ from
(2.13). This is a general  feature, the weight associated to the
singled-out term for N legs of type A is ${\cal
N}(q_1,\ldots,q_N)(-1)^{N-1}$ as in (2.13). If one would disregard
all discontinuities, only this term would survive, and one would
get the form obtained in ref.[5] for bare vertices with any number
of legs A and R.

 This is an example of the general validity of the
relation between RA and IT amplitudes.

\noindent As for the coefficient multiplying the discontinuity
in $q+s$, its role is to symmetrize the weight among $r$ , $t$
and the particles in the intermediate state in the $(q+s)$
channel.
One property of the R,A basis is emerging, it insists on a
symmetric weight with respect to all A legs.

\bigskip

{\large \bf 3.3 \ The general case}

\bigskip
The general case with two A and any number $s$ of R

follows. There are only two ways of splitting the
$\epsilon$-flow, (one sink-one source) and (one sink-one
source-one sink). On the diagrams drawn on fig.5 the additional
legs may be attached to the right blob on diagram (a), and to
both side-blobs on diagram (b). All the possible cases for the
$\epsilon$ variables are exhausted for

$\epsilon_p+\epsilon_q+\sum_{m=1}^s\epsilon_m=0$

with $\epsilon_m>0$
 and $\epsilon_p , \epsilon_q\ <0$ .
\\

Indeed if $\epsilon_p+\epsilon_i>0$ then
$\epsilon_q+\sum_{m\not=i} \epsilon_m<0$ for any sum of subsets
of $\epsilon_m$. If $\epsilon_p+\epsilon_i<0$ , one considers
$\epsilon_p+\epsilon_i+\epsilon_j$ and repeats the argument. The
convenient energy variables (that are undetermined) are the
energy component of the multiperipheral-momentum-transfer
variables, familiar at $T=0$ [9], i.e. $p^0+\sum_{m=1}^k p_m^0$.
For a given sequence of such variables $p+p_i$ , $p+p_i+p_j$ ,
\ldots, the sum rule that generalizes (3.6) is

\begin{equation}
N(p,q+\sum_{m\not=i}p_m)+N(p+p_i,q+\sum_{m\not=i,j}p_m)+
\cdots+N(p+\sum_{m\not=s}p_m,q)=N(p,q)
\end{equation}
Each amplitude differs from its neighbour by a discontinuity in
one single variable. Then, the general relation between RA and IT
amplitudes involves one analytical continuation with weight
$N(p,q)$ and discontinuities in an individual momentum-transfer
variable.

The general case with three A and any number of R contains the
following way of splitting the $\epsilon$-flow: (one
source-one sink), (one sink-one source-one sink) , (one
source-one sink-one source-one sink) , (one sink-one source-one
sink-one source-one sink) . The convenient energy variables are
the  momentum transfer variables, and the sum rule for the
coefficients generalizes (3.10).

\bigskip
To conclude, in this sect.3 the relations between RA and IT
amplitudes have been explicitely given for the case when more than
one leg differs from the other ones' type.
They are (3.4) for $N=4$-point functions and (3.8) for
$N=5$-point functions. The coefficients of the linear relations are
statistical weights and they obey a sum rule. The RA amplitude is
obtained as the sum of one analytical continuation of the IT
amplitude plus discontinuities in individual variables. The
multiperipheral momentum transfers appear as energy variables. Two
discontinuities are involved for $N=4$, six enter for $N=5$.

\bigskip

\bigskip

\begin{center}
{\Large \bf 4. \ Rules for the discontinuities}
\end{center}
\setcounter{chapter}{4}
\setcounter{equation}{0}

\bigskip

\bigskip
 The relation between RA and IT amplitudes generally
involves the discontinuity of IT amplitudes across cuts in a given
channel. Such a discontinuity obeys Cutkosky rules where the
intermediate states are the familiar ones at $T=0$. All
the signs are allowed for the intermediate-particles' energies.
The results of this section are direct consequences of the IT
amplitudes' form obtained in ref.[7] as a sum over a set of
intermediate states, with a factorization property. In this
work  an alternate
 proof will be given in terms of the $\epsilon$-flow.
 It is valid for any diagram. The evaluation of a
discontinuity across a cut in a single momentum-transfer variable
is very similar to the self-energy case. The  method will be
introduced for the self-energy, then applied to the
discontinuities of interest.

\bigskip

{\large \bf 4.1 \  The self-energy case}

\bigskip
\noindent For a diagram contributing to the self-energy , the
discontinuity is given by (see fig.6)

\begin{equation}
\Sigma(p_0+i\epsilon)-\Sigma(p_0-i\epsilon)=\Gamma_{R\ \ A}^{p,-p}
-\Gamma_{A\ \ R}^{p,-p}=\sum_m{\cal N}(q_1,\ldots,q_m)\
\ I_m^{(1)}
\end{equation}
\begin{equation}
I_m^{(1)}=\Gamma^{m+1}(p_A,q_{1R},\ldots,q_{mR})\ \ \Gamma^{m+1}
((-p)_A,(-q_1)_R,\ldots,(-q_m)_R)
\end{equation}
\begin{equation}
I_m^{(1)}=\Gamma^{m+1}(p_R,q_{1A},\ldots,q_{mA})\ \ \Gamma^{m+1}
((-p)_R,(-q_1)_A,\ldots,(-q_m)_A)
\end{equation}
where the $\Gamma^{m+1}$ are the $(m+1)$-point IT function
introduced in (2.12) and (2.13), one on each side of the
cutting plane. ${\cal N}(q_1,\ldots,q_m)$ is the statistical
weight of an $m$-particle intermediate state defined in
(2.11), and the sum is over all the possible intermediate
states of the diagram
\begin{equation}
\sum_m=2\pi\sum_m\int\prod_{i=1}^{m}({\rm d}_4q_i\
\epsilon(q_i^0)\ \delta({\underline q}_i^2-m^2))
\end{equation}

\medskip
The basic tool for the graphical proof  follows. The
difference between the two orientations of the
$\epsilon$-flow along a string of propagators $BC\ldots L$
 may be written
\begin{equation}
 B_R-B_A=\Delta_R({\underline
k}_B)-\Delta_A({\underline k}_B)=2\pi\ \epsilon(k_B^0)\
\delta({\underline k}_B^2-m^2)

\end{equation}
\begin{equation}
B_RC_R-B_AC_A=(B_R-B_A)C_R+B_A(C_R-C_A)=B_R(C_R-C_A)+(B_R-B_A)C_A
\end{equation}
\begin{eqnarray}
B_RC_RD_R-B_AC_AD_A&=&B_RC_R(D_R-D_A)+B_R(C_R-C_A)D_A+
(B_R-B_A)C_AD_A   \nonumber \\ &=&
(B_R-B_A)C_RD_R+B_A(C_R-C_A)D_R+B_AC_A(D_R-D_A)

\end{eqnarray}
The general rule is, the difference is the sum over all possible
ways of cutting the string $BC\ldots L$ into two pieces. To the cut
line is associated the factor (4.5) (no thermal factor), and the
$\epsilon$-flow may be directed either away from the cut line in
all terms, or towards the cut line.

\noindent The $method$ for the proof is now described. The IT
amplitude for a loop diagram contributing to the self-energy may be
written as a sum of trees. The $\epsilon$-flow is from left
to right in $\Gamma_{R\ A}^{p,-p}$\ , from right to
left in $\Gamma_{A\ R}^{p,-p}$ . In the difference
between the two cases, one internal line is necessary on
shell (from the above rule) and each tree is split into
two pieces.  Some examples will illustrate the way one obtains the
weight ${\cal N}(q_1,\ldots,q_m)$ of the intermediate state and
the Green functions on each side of the cutting plane.

\medskip
The IT amplitude for the one-loop diagram is the sum of two
trees, each with a factor $(n(q_i^0)+1/2) \epsilon(q_i^0)$
attached to the cut line $i$, according to the rule of sect.2.1.
See fig.7(a). In the difference between the two orientations of
the $\epsilon$-flow, the remaining internal line is on shell, from
(4.5), with no thermal factor. Each tree is split into two pieces,
which are identical for both trees. The weight
$N(q_1,q_2)$ is obtained as the sum of the two cases

$$2\pi\epsilon(q_1^0)\delta({\underline
q}_1^2-m^2)\epsilon(q_2^0)\delta({\underline q}_2^2-m^2)
(n(q_1^0)+1/2\ +\ n(q_2^0)+1/2)$$ i.e. form (4.1) to (4.3) holds
for that case.

\noindent As a two-loop example, one considers the tree diagram
drawn on fig.7(c), where lines $a$ and $c$ are cut. Three terms
are generated in the difference between the two orientations of the
$\epsilon$-flow along the string of propagators ( $b\ e\ d$ ): i)
line $e$ is on shell, the term is one of the three needed for the
intermediate state ( a e c ) ii) line $b$ is on shell, the terms
is one of the six needed for the intermediate state ( a b ) \ (the
one-loop vertex will appear as the sum of three trees) \ iii) line
$d$ is on shell, the term contributes to the intermediate state (
c d ). The same analysis may be done for the seven others trees
associated to the diagram drawn on fig.7(b). It is a matter of
book-keeping to group the terms according to the four possible
intermediate states of the diagram ( a b ) ( c d ) ( b e d ) (a e
c ).

\bigskip
The general argument follows.

\noindent For an arbitrary diagram, one looks at a
possible intermediate state of the diagram, say a
cutting plane goes through the lines $1,2,\ldots,m$ .
When the diagram is written as a sum of trees, one
considers the subset of trees where those $m$ lines are
on shell, except one, say line $j$ . Line $j$ carries the
energy $p_0+\sum_{i=1,i\not=j}^m q_i^0$ . The difference
between the two orientations of the $\epsilon$-flow from
the external line $p$ to the external line $-p$
generate terms and one term corresponds to line $j$
on-shell (with no thermal factor). In this case each tree of the
subset is split into two pieces. Collecting all those
pieces one builds up, on each side  of the cutting
plane, a Green function, written as a sum of trees.
{}From (4.5) and the rule of sect.2.1, the Green functions are
multiplied by a factor

\begin{equation}
W_j=2\pi \prod_{i=1}^m(\epsilon(q_i^0)\delta({\underline
q}_i^2-m^2))\ \
\prod_{i=1,i\not=j}^m(n(q_i^0)+{1\over2})
\end{equation}
As the Green functions are the same for any $j$, the
total weight associated to the intermediate state is
\begin{equation}
W=2\pi \prod_{i=1}^m(\epsilon(q_i^0)\delta({\underline
q}_i^2-m^2))\ \
\sum_{j=1}^m \ \ \prod_{i=1,i\not=j}^m(n(q_i^0)+{1\over2})
\end{equation}
The full weight ${\cal N}(q_1,\ldots,q_m)$ as given by (2.11) may
be written
  \begin{eqnarray}
{\cal N}(q_1,\ldots,q_m)&=&\sum_{j=1}^m \ \

\prod_{i=1,i\not=j}^m(n(q_i^0)+{1\over2})+{1\over4}\sum_{j,k,l}
 \ \ \prod_{i\not=j,k,l}(n(q_i^0)+{1\over2})
\nonumber \\ & &
+\mbox{}\cdots+{1\over{2}^m}(1-(-1)^m)
\end{eqnarray}
with the use of $n+1=n+1/2+1/2  \ \ ,\ \
\ n=n+1/2-1/2$ . Comparing

the  factor in (4.9) to (4.10), one finds that the
lowest order terms in the B.E. factor are missing. The reason is,
some trees should have lowest order terms in the numerator in
addition to $\prod_{i\not=j}^m(n(q_i^0)+1/2)$ . The rules to
write down those extra terms in some trees were given
up to the fourth loop order in ref.[7], where they were
shown to generate the full weight
${\cal N}(q_1,\ldots,q_m)$ of the intermediate state.

As for the $\epsilon$-flow, there is a choice, either
away or towards the cut line, from (4.7). The same choice in all
 trees guaranties that the sum of the oriented
pieces of trees is the Green function where the
external momentum $p$ is the single sink (or source).
Forms (4.2) and (4.3) are then obtained.

\bigskip

{\large \bf 4.2 \  The discontinuity in a given channel}

\bigskip
In sect.3 the relation between $\Gamma _{AARR}^{ \ pqsr}$ and the
IT amplitudes was obtained. It involves the difference
between two analytical continuations of the IT
amplitudes differing only in one two-particle energy.
For example in (3.4) it appears
\begin{equation}
\delta \Gamma=\Gamma^4((p+r)_R,(p+s)_A)-\Gamma^4((p+r)_A,(p+s)_A))
\end{equation}
When the IT amplitude associated to a diagram is written as a sum
of trees, only  contribute to (4.11) the trees where the momentum
$p+r$ is carried by a string of propagators . When going  from
$(p+r)_A$ to $(p+r)_R$ in (4.11), only the flow along the lines
carrying this momentum  is reversed. In the difference between the
two cases, each internal line carrying the momentum $p+r$ is
successively on shell. One then considers the possible
intermediate states  of the full diagram in the channel $(p+r)$
and the general argument goes as for the self-energy case.
 One obtains (see
fig.8) \begin{eqnarray}
&&{\Gamma^4(p_A,q_A,r_R,s_R,(p+r)_R,(p+s)_A)-
\Gamma^4(p_A,q_A,r_R,s_R,(p+r)_A,(p+s)_A) \hspace*{\fill} }
\nonumber  \\  && =\sum_m
{\cal N}(q_1,\ldots,q_m) \ \ I_m^{(2)}

\end{eqnarray}
\begin{equation}
I_m^{(2)}=\Gamma^{m+2}(p_A,r_R,q_{1R},\ldots,q_{mR})\
\ \Gamma^{m+2}(q_A,s_R,(-q_1)_R,\ldots,(-q_m)_R)
\end{equation}
\begin{equation}
I_m^{(2)}=\Gamma^{m+2}(p_A,r_R,q_{1A}\ldots,q_{mA})\ \
\Gamma^{m+2}(q_A,s_R,(-q_1)_A,\ldots,(-q_m)_A)
\end{equation}
where the sum is over all possible intermediate states
in the $(p+r)$ channel. Both choices are possible for
the $\epsilon$-flow, away or towards the cutting plane,
as there is one sink or source in each piece in both
cases.

 For the 5-point function $\Gamma_{AARRR}^{ \ pqrst}$, the
same type of discontinuity of the IT amplitude in
a single sub-energy variable appears in (3.8). For example, one
looks at  the discontinuity  in $p+r$. In the subset of trees
depending on the pair of variables $(\ p+r\ ,\ q+s\ )$, discussed
in sect.3.2.1, only the lines carrying the momentum $p+r$ will be
successively on shell when one computes the discontinuity
\begin{eqnarray}
\lefteqn{\Gamma^5((p+r)_R,others A)-\Gamma^5((p+r)_A,others A)=}
\nonumber \\ &&
\sum_m{\cal N}(q_1,\ldots,q_m)\
\Gamma^{m+2}(p_A,r_R,q_{1R},\ldots,q_{mR})
 \ \Gamma^{m+3}(q_A,t_R,s_R,(-q_1)_R,\ldots,(-q_m)_R)
  \end{eqnarray}
where the sum is over all possible intermediate states in the
channel $(p+r)$. Contrary to the previous cases, only one form
exists. Indeed the result must be written in terms of a single
source or sink in each piece.

The property is valid generally. For the N-point functions
considered in sect.3.3, reversing the flow in only one
multiperipheral-momentum-transfer energy variable, keeping the
other fixed, will isolate the intermediate states for that
channel.

\bigskip
The conclusion of this section is, a discontinuity in an
individual momentum-transfer energy-variable is associated to
processes where particles are on shell in a possible intermediate
state for that channel. The corresponding term factorizes into a
thermal statistical weight (of the multiparticle intermediate
state) and two IT amplitudes with a single sink (or source) for
the $\epsilon$-flow in each one.

\bigskip

\bigskip

\begin{center}
{\Large \bf 5. The real parts of the 3-point function}
\end{center}
\setcounter{chapter}{5}
\setcounter{equation}{0}

\bigskip

\bigskip
 To a 3-point function in the R,A basis corresponds one
analytical continuation of the IT amplitude. The functions
$\Gamma_{RRA}$ , $\Gamma_{ARR}$ , $\Gamma_{RAR}$ are different
analytical continuations of the IT amplitude. One wish to know
how their real parts differ from each other, in connection with
the definition of an effective coupling constant.

 In this section, a general formula will be obtained for the
difference between two real parts. It involves four analytical
continuations and it amounts to taking twice a discontinuity.
The proof will use the same method in terms of the
$\epsilon$-flow as the one used in sect.4.1 for the self-energy.
\bigskip

Following ref.[8], one gets rid of the weight $N(p,s)$ on the pair
of legs A

\begin{equation}
\Gamma_{AAR}^{\ pst}=-N(p,s)\ \ {\tilde{\Gamma}}_{AAR}^{\ pst}
\end{equation}
so that relation (2.16) becomes
\begin{equation}
{\tilde{\Gamma}}_{AAR}^{\ pst}=\Gamma_{RRA}^{\ pst \ \ast}
\end{equation}
and one considers the quantity $\Gamma_{RAR}^{\ pst}-
\tilde{\Gamma}_{AAR}^{\ pst}$ . With

$\epsilon_p+\epsilon_s+\epsilon_t
=0$ and $\epsilon_s<0 \ ,\ \epsilon_t>0$ ,
\\
 one may have
$\epsilon_p>0$ or $\epsilon_p<0$. A diagram contributing to the
3-point function possesses intermediate states in the $p$
channel, the $s$ channel, the $t$ channel. Keeping

$\epsilon_s,\epsilon_t$ fixed, the above difference selects the
intermediate states in the $p$ channel. The following relation
holds
\begin{equation}
\Gamma_{RAR}^{\ pst}-\tilde{\Gamma}_{AAR}^{\ pst}=\sum_m\ {\cal
N}(q_1,\ldots,q_m)\ \ \ I_m^{(3)}
\end{equation}
\begin{equation}
I_m^{(3)}=\Gamma^{m+1}(p_A,(-q_1)_R,\ldots,,(-q_m)_R)
\ \ \Gamma^{m+2}(s_A,t_R,q_{1R},\ldots,q_{mR})
\end{equation}
\begin{equation}
I_m^{(3)}=\Gamma^{m+1}(p_R,(-q_1)_A,\ldots,(-q_m)_A)\ \

\Gamma^{m+2}(s_A,t_R,q_{1A},\ldots,q_{mA})
\end{equation}
The sum runs over all the possible intermediate states of the
diagram in the $p$ channel. Note that, in terms of the
$\epsilon$-flow, there is one single sink (or source) in each
Green function of (5.4) or (5.5).

\bigskip
 One and two-loop examples
are now given with a graphical proof in terms of the
$\epsilon$-flow. When passing from $\Gamma_{RAR}^{pst}$ to
$\tilde{\Gamma}_{AAR}^{pst}$ , one goes from one sink  $(s)$  to
one source  $(t)$ . The $\epsilon$-flow along the lines that
connect $t$ to $s$ is unchanged, the flow along the other ones is
reversed.

\noindent For one tree of the one-loop diagram, the difference
between the two cases is drawn on fig.9. Adding the tree where
the roles of lines $a$ and $b$ are interchanged, one obtains the

weight of the
intermediate state ( a b ) in
a way similar to the one-loop self-energy case drawn on fig.7.
The third tree disappears, and forms (5.3) to (5.5) are valid
for that case.

\noindent One tree of a two-loop diagram is shown on fig.10(b)
where  lines $b$ and $d$ are cut. The change from one sink $(s)$ to
one source $(t)$ reverses the flow along the string of internal
lines ( $a\ e\ c$ ). The term with $a$ on shell contributes to the
intermediate state ( a b ), the term with $e$ on shell to ( b e d
), the term with $c$ on shell to ( c d ), in a way similar to the
two-loop self-energy diagram discussed previously in sect.4.1.

\noindent There are ten other trees associated to the
diagram shown on fig.10(a). One just has
to group the terms according to the intermediate states in
the $p$ channel  : ( a b ), ( c d ), ( b e d ), ( a e c ).
Forms (5.4) and (5.5) correspond to the two options for the
$\epsilon$ flow, either away or towards the cut line.

\bigskip

\bigskip
One now considers the case where the roles of legs $s$ and $t$
are interchanged and  one substracts both cases
\begin{equation}
2X=(\Gamma_{RAR}^{\ pst}-\tilde{\Gamma}_{AAR}^{\ pst})-
(\Gamma_{RRA}^{\ pst}-\tilde{\Gamma}_{ARA}^{\ pst})
 \end{equation}
{}From (5.2) $X$ is a real quantity
\begin{equation}
 2X=\Gamma_{RAR}^{\ pst}+\Gamma_{RAR}^{\ pst\
\ast}-(\Gamma_{RRA}^{\ pst}+\Gamma_{RRA}^{\ pst\ \ast})
\end{equation}
{}From (5.6),(5.3)(5.4) $X$ is

\begin{equation}
2X=\sum_m\ {\cal N}(q_1,\ldots,q_m)\ \
\Gamma^{m+1}(p_A,(-q_1)_R,\ldots,(-q_m)_R)\ \ \ I_m^{(4)}
\end{equation}
\begin{equation}
I_m^{(4)}=\Gamma^{m+2}(s_A,t_R,q_{1R},\ldots,q_{mR})-
\Gamma^{m+2}(s_R,t_A,q_{1R},\ldots,q_{mR})
\end{equation}
With $\epsilon_s+\epsilon_t+\sum_j\epsilon_j=0$ and
$\epsilon_j>0 \ \ j=1,\ldots,m$ , the change from
$\epsilon_s<0\ \ \epsilon_t>0$ to $\epsilon_s>0\ \
\epsilon_t<0$ reverses the sign of

$\epsilon_s+\sum_{j\in
a}\epsilon_j=-(\epsilon_t+\sum_{j\in b}\epsilon_j)$
 \\
for any partition of $q_1,\ldots,q_m$ into two sets $a$ and
$b$, i.e. it changes the way one approaches the real axis
for all the energy variables $s^0 + \sum_{j\in a}q_j^0$. The
difference $I_m^{(4)}$ isolates the discontinuity in any of those
channels \begin{equation}
\Gamma^{m+2}(s_A,t_R,q_{1R},\ldots,q_{mR})-
\Gamma^{m+2}(s_R,t_A,q_{1R},\ldots,q_{mR})=\sum_n{\cal
N}(r_1,\ldots,r_n) \ \ K_{mn}
\end{equation}
\begin{equation}
K_{mn}=\Gamma(s_A,q_{1R},\ldots,q_{kR},r_{1R},\ldots,r_{nR})
\ \ \Gamma(t_A,q_{k+1 R},\ldots,q_{mR}, (-r_1)_R,\ldots,(-r_n)_R)
\end{equation}
The sum runs over all the possible intermediate states of the
diagram's $(m+2)$-point function

$\Gamma^{m+2}(s,t,q_1,\ldots,q_m)$ so that $s$ sits on one side
of the cutting plane and $t$ on the other side. Indeed a tree may
have simultaneous poles in the channels \ $s\ ,\ s+q_i\ ,
s+q_i+q_j\ ,\ldots,\ t$  and one may perform an expansion in
simple poles of it [7].

\noindent An alternate form to (5.8)(5.9)(5.10) is obtained from
(5.5) where the role of legs A and R are interchanged. Relations
similar to (5.3)(5.8) exist in the two other channels of the
3-point function.
\medskip

As an example,  the two-loop diagram of fig.10(a) is considered.
The interchange of $s$ and $t$ reverses the $\epsilon$-flow along
the lines that connect $s$ to $t$. For the intermediate state ( b
e d ) of the $p$ channel, the difference between the cases $s_A\
t_R$ and $s_R\ t_A$ puts line $f$ on shell. For the intermediate
state ( a b ) of the $p$ channel, the one-loop 4-point function
$(\ a\ b\ s\ t\ )$ is a sum of four trees. The difference
$\Gamma(s_A,t_R,a_R,b_R)-\Gamma(s_R,t_A,a_R,b_R)$ generates the
sum over the intermediate states ( f d ), ( f e ), ( f c ) . One
just draw the diagrams, performs the difference as previously
done, and arranges the terms
according to these intermediate states. The choice of the
$\epsilon$-flow away from the cut lines guaranties that there is
a single sink on each side of the cutting plane, i.e. all the
oriented pieces of trees sum up to one Green function.

\bigskip
To conclude this section, the general relation for the
difference between the two real parts is summarized by the diagram
drawn on fig.11. It emphasizes the fact that the real parts of
the amplitudes $\Gamma_{RAR}$ and $\Gamma_{RRA}$ differ from
each other by a double discontinuity. This quantity is the
analogue of the Mandelstam double spectral function for the
4-point function at $T=0$. That difference between the real
parts of the two analytical continuations $( \epsilon_p>0\ ,\
\epsilon_s<0\ ,\ \epsilon_t>0\ )$ and $( \epsilon_p>0\ ,\

\epsilon_s>0\ ,\ \epsilon_t<0\ )$ is expressed in terms of
three one-particle "decay" amplitudes. The relevance of those
processes is a manifestation of the plasma, as the
intermediate particles may be in an initial or final state.

\bigskip

\bigskip

{\Large \bf 6. \ Conclusion}

\bigskip

\bigskip
 A Retarded-Advanced N-point amplitude of the
Real-Time formalism has been expressed as a linear combination of
analytic continuations of the corresponding Imaginary Time
amplitude. The coefficients appearing in the linear
combination are statistical factors. They have been obtained
for bosons and to generalize to the case of bosons and
fermions is straightforward. The full set of relations are written
explicitely for N=3,4,5-point Functions.

\noindent Two examples of the general validity of these
relations have been encountered: for particular cases, one recovers
 the results obtained  with the R,A Feynman rules [4] or the
theory with multiparticle bare vertices is recovered [5].

\bigskip
In perturbation theory, the computation of a multi-loop diagram
is much simpler for the IT amplitude. This is because one does not
have to keep track of the double  degree of freedom. In
the R,A basis that doubling takes a simple form, the oriented
flow of  infinitesimal $\epsilon$. An interesting and general
feature of an IT amplitude associated to a diagram is that it may
be written as a sum of trees. The RA amplitudes are expressed in
terms of IT amplitudes where the flow along these trees is fixed.
Thus one has an explicit systematic form for an RA amplitude
associated to a diagram. One can then go from the usual (1,2)
basis to the (R,A) basis with the use of (U,V) matrices on the
external legs of the N-point amplitude.

\medskip
The RA amplitude has been obtained as the sum of one analytical
continuation of an IT amplitude plus discontinuities in
specific channels corresponding to momentum transfer variables.

Those discontinuities have been associated to process where
particles are on-shell in a possible multiparticle
intermediate state for that channel. The corresponding term
factorizes into a thermal statistical weight for the
intermediate particles and two IT amplitudes with a specific
analytical continuation.

\medskip
One physical picture that is recurrent is the presence of the
plasma. It can manifest itself in two ways

i) the $T$-dependance. The whole $T$-dependance of an RA or
IT amplitude is in the thermal statistical factors of

the multiparticle states.

ii) the role of on-shell particles in the intermediate
states. Their growing relevance comes from the fact that these
particles may be in the initial or final states, as plasma
particles. This extra kinematical freedom allows for the
presence of cuts that were forbidden in the "physical channel"
at $T=0$.

The several ways of approaching these extra cuts are responsible
for the complications that arise when relating RA and IT
amplitudes. The comparison between these two types of amplitudes
shows that the R,A basis puts an extra symmetric statistical
weight on all legs of type A when they are i) external momenta

ii) intermediate state momenta in a crossed channel.

\bigskip
Another interesting result of this work is in connection with the
definition of an effective coupling constant. It has been shown
that the real parts of two vertex functions  differ only by a
double discontinuity, which is expressed in terms of three
one-particle decay amplitudes.

\bigskip\bigskip\bigskip
\centerline{ \bf Acknowlegments}
\bigskip
I wish to thank P. Aurenche for his interest, and for bringing to
my attention the paper of J. C. Taylor.

\bigskip

\bigskip

\centerline{\large \bf Appendix}

\bigskip
One-loop diagrams with RA Feynman rules

\bigskip
The external momenta are incoming. The positive orientation
around the loop is the trigonometric one for both momentum and
$\epsilon$-flow. The notations are
$${\bf 1^+}=\Delta_R(k_1) \ ,\ {\bf 1^-}=\Delta_R(-k_1) \ ,\

\ 1^+=k_1\ \ ,1^-=-k_1$$
$$n_{1^+}=n(k_1^0)+1/2 \ ,\ n_{1^-}=n(-k_1^0)+1/2 \ ,\
n_{1^+}+n_{1^-}=0 $$

The 4-point one-loop diagram is chosen as an example. See fig.3
for the definition of the momenta. The conservation of momentum at
the vertices is

$$p+0^++1^-=0=q+3^++0^-=r+1^++2^-=s+2^++3^-$$

\bigskip
One considers this diagram in the case of one leg of type A. With
the use of the RA Feynman rules, the sum over all possible cases
for the $\epsilon$-flow around the loop is
\begin{eqnarray}
-\  \Gamma_{ARRR}^{\ pqst}&=&{\bf 0^+1^+2^+3^+}N(p,1^-)+
{\bf 0^-1^-2^-3^-}N(p,0^+) \nonumber \\ & &
+{\bf 0^+1^-}[N(1^+,2^-){\bf 2^+3^+}+N(2^+,3^-){\bf 2^-3^+}+
N(3^+,0^-){\bf 2^-3^-}] \nonumber
\end{eqnarray}
Terms such as ${\bf 0^+1^+2^+3^+} n_p$ integrate to zero (see
ref.[4]). $N$ is written as in (2.8) and the terms are arranged
to get
\begin{eqnarray}
-\  \Gamma_{ARRR}^{\ pqst}&=& (n_{1^-}{\bf 1^+}+n_{1^+}{\bf 1^-})

{\bf 2^+3^+0^+}+(n_{0^+}{\bf
0^-}+n_{0^-}{\bf 0^+}){\bf 1^-2^-3^-} \nonumber \\ & &
+(n_{2^-}{\bf 2^+}+n_{2^+}{\bf 2^-}){\bf 3^+0^+1^-}+
(n_{3^+}{\bf 3^-}+n_{3^-}{\bf 3^+}){\bf 0^+1^-2^-} \nonumber
\end{eqnarray}
with
 $$(n_{1^-}{\bf 1^+}+n_{1^+}{\bf 1^-})
=-n_{1^+}({\bf 1^+-1^-})=-(n(k_1^0)+1/2)2\pi
\epsilon(k_1^0)\delta(k_1^2-m^2) $$
from (2.4). $\Gamma_{ARRR}^{\ pqst}$ is now written as a sum of
oriented trees where the factor (2.1) is attached to the cut line.

\bigskip
The one-loop diagram is now computed in the case of two legs of
type A. With the use of RA Feynman rules, the result is
\begin{eqnarray}
\Gamma_{AARR}^{\ pqst}&=&{\bf 0^+1^+2^+3^+}N(p,1^-)N(q,0^-)
+{\bf 0^-1^-2^-3^-}N(p,0^+)N(q,1^+) \nonumber \\ & &
+{\bf 1^-3^+}[N(1^+,2^-){\bf 2^+}+N(2^+,3^-){\bf 2^-}]
[N(p,0^+){\bf 0^-}+N(q,0^-){\bf 0^+}] \nonumber
\end{eqnarray}
Form (2.10) is used for $N$ to obtain
$$N(p,1^-)N(q,0^-)=N(p,q)N(1^-,p+q)=N(p,q)[n_{1^-}+n_{p+q}]$$
$$N(1^+,2^-)N(p,0^+)=N(0^+,2^-)N(p,2^-+0^+)=(n_{0^+}+n_{2^-})
N(p,q+s)$$
\begin{eqnarray}
N(1^+,2^-)N(q,0^-)&=&n_qN(1^+,2^-)+n_pN(1^+,0^-)
-n_{0^++2^-}N(0^+,2^-) \nonumber \\ &=&
n_{1^+}N(p,q)+n_{2^-}N(q,0^-+2^+)+n_{0^-}N(p,0^++2^-) \nonumber
\end{eqnarray}
The term associated to $n_{p+q}$ integrates to zero and one gets
\begin{eqnarray}
\Gamma_{AARR}^{\ pqst}&=&[{\bf 2^+3^+0^+}(n_{1^+}{\bf
1^-}+n_{1^-}{\bf 1^+}) +(n_{3^-}{\bf 3^+}+n_{3^+}{\bf 3^-})
{\bf 2^-1^-0^-}]N(p,q) \nonumber \\ & &
+(n_{2^-}{\bf 2^+}+n_{2^+}{\bf 2^-}){\bf 1^-3^+}[{\bf 0^-}N(p,q+s)
+{\bf 0^+}N(q,p+r)] \nonumber \\ & &
+(n_{0^-}{\bf 0^+}+n_{0^+}{\bf 0^-}){\bf 1^-3^+}
[{\bf 2^+}N(p,q+s)+{\bf 2^-}N(q,p+r)] \nonumber
\end{eqnarray}
$\Gamma_{AARR}^{\ pqst}$ is written as a sum of oriented trees
where the extra factors $N$ are as in (3.2) with $F_2=0$.

\newpage
\centerline{\large \bf References}
\bigskip
[1] T. S. Evans, Nuclear Phys. {\bf B374} (1992) 340.

\medskip
[2] T. S. Evans, Phys. Lett. {\bf B252} (1990) 108; {\it ibid.}
 {\bf B249} (1990) 286.

\medskip
[3] R. Kobes, Phys.Rev. {\bf D42} (1990) 562; {\it ibid.} {\bf
D43} (1991) 1269.

\medskip
[4] P. Aurenche and T. Becherrawy, Nucl.Phys. {\bf B379} (1992)
259.

\medskip
[5] M. A. van Eijck and Ch. G. van Weert, Phys.Lett. {\bf B278}
(1992) 305.

\medskip
[6] J. C. Taylor, Phys.Rev. {\bf D47} (1993) 725.

\medskip
[7] F. Guerin, INLN 1991/11, to be published.

\medskip
[8] P. Aurenche, E. Petitgirard and T. del Rio Gatzelurrutia,
Phys.Lett. {\bf B297} (1992) 337.

\medskip
[9] E. Byckling and K. Kajantje, {\it Particle Kinematics} (J.

Wiley, London, 1973) p.166.

\newpage
\centerline{\large \bf Figure Captions}

\bigskip

\bigskip
Fig.1. \ \ Bare propagators and vertices in the R,A basis. All
momenta are incoming, the arrow describes the $\epsilon$-flow.

\bigskip

\bigskip
Fig.2. \ \ Tree diagrams for $\Gamma_{AARR}^{\ pqsr}$ . All momenta
are incoming, arrows are for the $\epsilon$-flow.

\bigskip

\bigskip
Fig.3. \ \ Trees associated to the one-loop diagram for
$\Gamma_{AARR}^{ \ pqsr}$.

\bigskip

\bigskip
Fig.4. \ \ The decomposition of $\Gamma_{AARR}^{ \ pqsr}$ in terms
of the $\epsilon$-flow. External momenta are incoming.

\bigskip

\bigskip
Fig.5. \ \ The possible cases for the $\epsilon$-flow when two

external legs are of type A.

\bigskip

\bigskip
Fig.6. \ \ The discontinuity for the self energy in terms of the
$\epsilon$-flow.

\bigskip

\bigskip
Fig.7. \ \ (a) The weight $N(q_1,q_2)$ of the two-particle
intermediate state is obtained as the sum of two terms.  A cut
line is a line on shell with a thermal factor, a line carrying a
cross is a line on shell with no thermal factor. \ (b) A two-loop
diagram for the self-energy. \ (c) One tree associated to diagram
(b).

\bigskip

\bigskip
Fig.8. \ \ The discontinuity in a momentum-transfer variable in
terms of the $\epsilon$-flow.

\bigskip

\bigskip
Fig.9. \ \ Reversing the flow of one external leg. Conventions
are those of fig.7(a).

\bigskip

\bigskip
Fig.10. \ \ (a) A two-loop diagram for the vertex \ (b) One tree
associated to diagram (a).

\bigskip

\bigskip
Fig.11. \ \ The difference $X$ between the real parts of
$\Gamma_{RAR}^{ \ pst}$ and $\Gamma_{RRA}^{ \ pst}$ as a double
discontinuity.

\bigskip

\bigskip

\end{document}